\documentclass[amsmath,amssymb,twocolumn,english]{revtex4}

\usepackage{graphicx}
\usepackage{dcolumn}
\usepackage{bm}
\usepackage[utf8]{inputenc}
\usepackage[T1]{fontenc}
\usepackage{mathptmx}
\usepackage{booktabs}
\usepackage{color}
\usepackage{multirow}

\begin{document}

\title{Benchmarking GNOF against FCI in challenging systems in one, two and three dimensions}

\author{Ion Mitxelena}     
\affiliation{Donostia International Physics Center (DIPC), 20018 Donostia, Euskadi, Spain; Euskal Herriko Unibertsitatea (UPV/EHU), PK 1072, 20080 Donostia, Euskadi, Spain}

\author{Mario Piris}
\email{mario.piris@ehu.eus}
\affiliation{Donostia International Physics Center (DIPC), 20018 Donostia, Euskadi, Spain; Euskal Herriko Unibertsitatea (UPV/EHU), PK 1072, 20080 Donostia, Euskadi, Spain; and Basque Foundation for Science (IKERBASQUE), 48009 Bilbao, Euskadi, Spain}

\date{\today}

\begin{abstract}
This work assess the reliability of the recently proposed [Phys. Rev. Lett. 127, 233001, 2021] global natural orbital functional (GNOF) in the treatment of the strong electron correlation regime. We first use an H$_{10}$ benchmark set of four hydrogen model systems of different dimensionalities and distinctive electronic structures: a 1D chain, a 2D ring, a 2D sheet, and a 3D close-packed pyramid. Second, we study two paradigmatic models for strongly correlated Mott insulators, namely a 1D H$_{50}$ chain and a 4x4x4 3D H cube. We show that GNOF without hybridization to other electronic structure methods and free of tuned parameters succeeds in treating weak and strong correlation in a more balanced way than the functionals that have preceded it.

\end{abstract}

\maketitle

\section{Introduction}
The vast majority of electronic structure methods are built onto a mean-field reference such as the Hartree-Fock wavefunction. While this approach is appropriate for most molecular systems, it fails to describe strongly correlated systems \cite{tew-jcc-2007}. The latter is also one of the main limitations of density functional theory (DFT) approximations \cite{cohen-dft-2012,wenna-jpcl-2022}, despite recent efforts to overcome the associated fractional electron problem \cite{science-2021-frac-dft}. Alternatively, reduced density matrix (RDM) methods hold the promise to give a complete description of the electron correlation \cite{mazziotti2012review,FossoTande2016,Piris2017,mazziotti2021,piris-prl-2021}. The second-order RDM (2RDM) $D$ alone is sufficient to describe many-electron systems in the non-relativistic limit, thus reducing dramatically the computational cost inherent to wavefunction expansions \cite{FossoTande2016}. In practical calculations, however, the optimization of the 2RDM must be subject to a set of N-representability conditions to assure, at least approximately, that the resulting 2RDM is derivable from an antisymmetrized N-electron density matrix \cite{mazziotti2012structure}. Two-particle N-representability conditions have been shown to lead to accurate results in many situations, and this accuracy can be improved by including three-particle constraints \cite{mazziotti2006,mazziotti2007}, but sometimes the latter are essential to describe some strongly correlated systems \cite{vertichel-prl-2012,rubin-mazziotti-tca-2014,li-eugene-deprince-jcp-2021}. Unfortunately, in the conventional formulation of the variational 2RDM (v2RDM) method \cite{mazziotti2012review}, the inclusion of three-particle conditions increases the computational cost to $O(\mathrm{M}^{9})$ and therefore limits its application to large systems, M being the dimension of the single-particle space. Recently \cite{mazziotti2016,mazziotti2020}, a new dual formulation of the v2RDM method has reduced to $O(\mathrm{M}^{6})$ the computational cost of implementing the T2 condition in both floating-point operations and memory storage, which provides a significant improvement in the calculation of strongly correlated systems with v2RDM. 

The use of the first-order RDM (1RDM) $\Gamma$ to describe the correlated electronic structure leads to a further reduction of the computational cost of 2RDM approaches, and do not compromise the accuracy for studying strongly correlated systems \cite{mitxelena2020efficient1,mitxelena2020efficient2}. In contrast to DFT, the one-electron energy functional is exactly known in terms of $\Gamma$, however, we have to settle for making approximations for the reconstruction of the unknown electron-electron interaction energy functional. In Ref. \cite{Piris2006}, a bottom-up approach was derived to impose N-representability constraints on the construction of the functional, rather than doing it along the optimization process as it is done in v2RDM methods. Enforcement of two-particle positivity conditions, together with the electron-pairing approach, lead us to PNOF5 \cite{piris2011-pnof5,Piris2013e}, a natural orbital functional (NOF) that retrieves the Löwdin-Shull functional \cite{lowdin:56pr} for singlet states in two-electron systems. Note that we refer to NOF since, in most applications, the spectral decomposition of the 1RDM is used to express it in terms of the natural orbitals and their occupation numbers. PNOF5 constitutes the first N-representable functional that includes electron correlation, since it is equivalent to an antisymmetrized product of strongly orthogonal geminals \cite{Pernal2013}. Accordingly, PNOF5 is able to provide an accurate prediction for the dissociation of different chemical bonds, including multiple-bond dissociations \cite{matxain2011-pnof5}, by providing the correct number of electrons at the dissociation limit, refuting the statement about 1RDM functionals recently made by Wenna et al. \cite{wenna-jpcl-2022}.

Strongly correlated systems are particularly challenging for any electronic structure method including the independent pair model PNOF5 \cite{mitxelena2017performance,mitxelena2018a}. Recently \cite{Piris2017,mitxelena2018a}, the incorporation of electron correlation between pairs has resulted in an efficient method for strongly correlated systems in one and two dimensions \cite{mitxelena2020efficient1,mitxelena2020efficient2}. The resulting NOF approximation, PNOF7, correlates the motion of the electrons in different pairs with parallel and opposite spins explicitly considering particle-hole symmetry, in the line of the original formulation of Bardeen, Cooper and Schrieffer (BCS) \cite{Superconductivity1}. PNOF7  outperforms \cite{mitxelena2017performance,Mitxelena_2018-corrig} the most common 1RDM approximations in the description of the one-dimensional (1D) Hubbard model and its variant Aubry-André model at half-filling. More importantly, PNOF7 and v2RDM with two- and three-particle PQGT' constraints are comparable to study the two-dimensional (2D) Hubbard model at half-filling, for which they approach reference auxiliary-field quantum Monte Carlo results. Nevertheless, only the former retains its accuracy at lower filling situations \cite{mitxelena2020efficient2}. In summary, PNOF7 compares with the state-of-the-art methods for describing the Hubbard model, regardless of the dimensionality and size of the system. However, in the case of hydrogen (H) model systems, which resemble the Hubbard model with long-range interactions, the performance of PNOF7 is not as accurate as for the Hubbard model. The accurate description of these models by a NOF is thereby the motivation of the present work.

H clusters combined with a minimal basis add a subtle delocalization to the Hubbard Hamiltonian and represent a challenging strong correlation problem \cite{mazziotti2010,motta-prx-2017,stair-evangelista-jcp-2020,li-eugene-deprince-jcp-2021}. In view of the difficulties shown by PNOF7 to describe the dissociation of the 2D H lattice \cite{mitxelena2020efficient2} and asymmetric dissociation of linear H chains \cite{mitxelena2020efficient1}, a functional form including interpair dynamical correlation effects could be more appropriate to the study of such systems. In this vein, the recently proposed global NOF (GNOF) \cite{piris-prl-2021} has demonstrated a balanced treatment of electron correlation effects in molecular systems with different spins, including full dissociation curves, and ionization potentials of first-row transition metal atoms. In this work, we first study the ability of GNOF to deal with the benchmark set of 1D-, 2D-, and three-dimensional- (3D) H systems employed in Refs. \cite{li-eugene-deprince-jcp-2021} and \cite{stair-evangelista-jcp-2020}. Second, we analyze the GNOF results on two paradigmatic models for strongly correlated Mott insulators, namely a 1D H$_{50}$ chain and a 4x4x4 3D hydrogen cube. We demonstrate that a pure NOF without hybridization to other electronic structure methods and free of tuned parameters is adequate to provide a complete description of strongly correlated systems.

\section{Global Natural Orbital Functional}\label{sec:theory}

The nonrelativistic Hamiltonian under consideration is spin coordinate free; therefore, a state with total spin $S$ is a multiplet, i.e., a mixed quantum state that allows all possible $S_z$ values. Next, we briefly describe GNOF for spin-multiplets. A more detailed description of this JKL-only NOF can be found in Ref. \cite{piris-prl-2021}.

We consider $\mathrm{N_{I}}$ single electrons which determine the
spin $S$ of the system, and the rest of electrons ($\mathrm{N_{II}}=\mathrm{N-N_{I}}$)
are spin-paired, so that all spins corresponding to $\mathrm{N_{II}}$
electrons altogether provide a zero spin. We focus on the mixed state of highest
multiplicity: $2S+1=\mathrm{N_{I}}+1,\,S=\mathrm{N_{I}}/2$. In the absence of single electrons ($\mathrm{N_{I}}=0$), the energy obviously reduces to a NOF that describes
singlet states.

For an ensemble of pure states $\left\{ \left|SM_{s}\right\rangle \right\} $,
we note that the expected value of $\hat{S}_{z}$ for the whole ensemble is zero. Consequently, the spin-restricted theory can be adopted even if the total spin of the system is not zero. We use a single set of orbitals
for $\alpha$ and $\beta$ spins. All the spatial orbitals will be then doubly occupied in the ensemble, so that occupancies for particles with $\alpha$ and $\beta$ spins are equal: $n_{p}^{\alpha}=n_{p}^{\beta}=n_{p}.$

Next, we divide the orbital space $\Omega$ into two subspaces:
$\Omega=\Omega_{\mathrm{I}}\oplus\Omega_{\mathrm{II}}$. $\Omega_{\mathrm{II}}$
is composed of $\mathrm{N_{II}}/2$ mutually disjoint subspaces $\Omega{}_{g}$.
Each of which contains one orbital $\left|g\right\rangle $ with $g\leq\mathrm{N_{II}}/2$,
and $\mathrm{N}_{g}$ orbitals $\left|p\right\rangle $ with $p>\mathrm{N_{II}}/2$,
namely, 
\begin{equation}
\Omega{}_{g}=\left\{ \left|g\right\rangle ,\left|p_{1}\right\rangle ,\left|p_{2}\right\rangle ,...,\left|p_{\mathrm{N}_{g}}\right\rangle \right\} .\label{OmegaG}
\end{equation}

Taking into account the spin, the total occupancy for a given subspace $\Omega{}_{g}$ is 2, which is reflected in the following sum rule: 
\begin{equation}
\sum_{p\in\Omega_{\mathrm{II}}}n_{p}=n_{g}+\sum_{i=1}^{\mathrm{N}_{g}}n_{p_{i}}=1,\quad g=1,2,...,\frac{\mathrm{N_{II}}}{2}. \label{sum1}
\end{equation}
Here, the notation $p\in\Omega_{\mathrm{II}}$ represents all the
indexes of $\left|p\right\rangle $ orbitals belonging to $\Omega_{\mathrm{II}}$. In general, $\mathrm{N}_{g}$ may be different for each subspace. In this work, $\mathrm{N}_{g}$ is equal to a fixed number for all subspaces $\Omega{}_{g}\in\Omega_{\mathrm{II}}$. We adopt the maximum possible value of $\mathrm{N}_{g}$ which is determined by the basis set used in calculations. From (\ref{sum1}), it follows that 
\begin{equation}
2\sum_{p\in\Omega_{\mathrm{II}}}n_{p}=2\sum_{g=1}^{\mathrm{N_{II}}/2}\left(n_{g}+\sum_{i=1}^{\mathrm{N}_{g}}n_{p_{i}}\right)=\mathrm{N_{II}}.\label{sumNpII}
\end{equation}
Similarly, $\Omega_{\mathrm{I}}$ is composed of $\mathrm{N_{I}}$
mutually disjoint subspaces $\Omega{}_{g}$. In contrast to $\Omega_{\mathrm{II}}$,
each subspace $\Omega{}_{g}\in\Omega_{\mathrm{I}}$ contains only
one orbital $g$ with $2n_{g}=1$. It is worth noting that each orbital
is completely occupied individually, but we do not know whether the
electron has $\alpha$ or $\beta$ spin: $n_{g}^{\alpha}=n_{g}^{\beta}=n_{g}=1/2$.
It follows that 
\begin{equation}
2\sum_{p\in\Omega_{\mathrm{I}}}n_{p}=2\sum_{g=\mathrm{N_{II}}/2+1}^{\mathrm{N_{\Omega}}}n_{g}=\mathrm{N_{I}}.\label{sumNpI}
\end{equation}
In Eq. (\ref{sumNpI}), $\mathrm{\mathrm{N}_{\Omega}=}\mathrm{N_{II}}/2+\mathrm{N_{I}}$
denotes the total number of suspaces in $\Omega$. Taking into account
Eqs. (\ref{sumNpII}) and (\ref{sumNpI}), the trace of the 1RDM is
verified equal to the number of electrons: 
\begin{equation}
2\sum_{p\in\Omega}n_{p}=2\sum_{p\in\Omega_{\mathrm{II}}}n_{p}+2\sum_{p\in\Omega_{\mathrm{I}}}n_{p}=\mathrm{N_{II}}+\mathrm{N_{I}}=\mathrm{\mathrm{N}}.\label{norm}
\end{equation}
Using ensemble N-representability conditions, we can generate a reconstruction functional for the 2RDM in terms of the occupation numbers that leads to GNOF:
\begin{equation}
E=E^{intra}+E_{HF}^{inter}+E_{sta}^{inter}+E_{dyn}^{inter}
\end{equation}
The intra-pair component is formed by the sum of the energies of the pairs of electrons with opposite spins and the single-electron energies of the unpaired electrons, namely
\begin{equation}
E^{intra}=\sum\limits _{g=1}^{\mathrm{N_{II}}/2}E_{g}+{\displaystyle \sum_{g=\mathrm{N_{II}}/2+1}^{\mathrm{N}_{\Omega}}}H_{gg}
\end{equation}
\begin{equation}
E_{g}=\sum\limits _{p\in\Omega_{g}}n_{p}(2H_{pp}+J_{pp}) 
+ \sum\limits _{q,p\in\Omega_{g},p\neq q}\Pi \left(n_{q},n_{p} \right)L_{pq}
\end{equation}
\noindent where
\begin{equation}
\Pi\left(n_{q},n_{p}\right) = \sqrt{n_{q}n_{p}}\left(\delta_{q\Omega^{a}}
\delta_{p\Omega^{a}}-\delta_{qg}-\delta_{pg}\right)
\end{equation}
\noindent and  $H_{pp}$ are the diagonal one-electron matrix elements of the kinetic energy and external potential operators. $J_{pq}=\left\langle pq|pq\right\rangle $ and $L_{pq}=\left\langle pp|qq\right\rangle $ are the Coulomb and exchange-time-inversion integrals, respectively. $\Omega^{a}$ denotes the subspace composed of orbitals above the level $\mathrm{N}_{\Omega}$ ($p>\mathrm{N}_{\Omega}$). The inter-pair Hartree-Fock (HF) term is
\begin{equation}
E_{HF}^{inter}=\sum\limits _{p,q=1}^{\mathrm{N}_{B}}\,'\,n_{q}n_{p}\left(2J_{pq}-K_{pq}\right)
\end{equation}
\noindent where $K_{pq}=\left\langle pq|qp\right\rangle$ are the exchange integrals. The prime in the summation indicates that only the inter-subspace terms are taking into account ($p\in\Omega{}_{f},q\in\Omega{}_{g},f\neq g$). $\mathrm{N}_{B}$ represents the number of basis functions considered. The inter-pair static component is written as
\begin{equation}
\begin{array}{c}
E_{sta}^{inter}=-\left({\displaystyle \sum_{p=1}^{\mathrm{N}_{\Omega}}\sum_{q=\mathrm{N}_{\Omega}+1}^{\mathrm{N}_{B}}+\sum_{p=\mathrm{N}_{\Omega}+1}^{\mathrm{N}_{B}}\sum_{q=1}^{\mathrm{N}_{\Omega}}}\right.
\left.{\displaystyle +\sum_{p,q=\mathrm{N}_{\Omega}+1}^{\mathrm{N}_{B}}}\right)' 
\Phi_{q}\Phi_{p} \\ \\ L_{pq} - \:\dfrac{1}{2}\left({\displaystyle \sum\limits _{p=1}^{\mathrm{N_{II}}/2}\sum_{q=\mathrm{N_{II}}/2+1}^{\mathrm{N}_{\Omega}}+\sum_{p=\mathrm{N_{II}}/2+1}^{\mathrm{N}_{\Omega}}\sum\limits _{q=1}^{\mathrm{N_{II}}/2}}\right)' \Phi_{q}\Phi_{p}L_{pq} \\ \\
{\displaystyle \:-\:\dfrac{1}{4}\sum_{p,q=\mathrm{N_{II}}/2+1}^{\mathrm{N}_{\Omega}}}K_{pq}
\end{array}
\end{equation}
\noindent where $\Phi_{p}=\sqrt{n_{p}h_{p}}$ with the hole $h_{p}=1-n_{p}$. Note that $\Phi_{p}$ has significant values only when the occupation number $n_p$ differs substantially from 1 and 0. Finally, the inter-pair dynamic energy can be conveniently expressed as 
\begin{equation}
\begin{array}{c}
E_{dyn}^{inter}=\sum\limits _{p,q=1}^{\mathrm{N}_{B}}\,'\,
\left[n_{q}^{d}n_{p}^{d} +\;\Pi\left(n_{q}^{d},n_{p}^{d}\right)\right]
\left(1-\delta_{q\Omega^{b}_{II}}\delta_{p\Omega^{b}_{II}}\right)L_{pq} \label{edyn}
\end{array}
\end{equation}

In Eq. (\ref{edyn}), $\Omega^{b}_{II}$ denotes the subspace composed of orbitals below the level $\mathrm{N_{II}}/2$ ($p\leq\mathrm{N_{II}}/2$), so interactions between orbitals belonging to $\Omega^{b}_{II}$ are excluded from $E_{dyn}^{inter}$. The dynamic part of the occupation number $n_{p}$ is defined as
\begin{equation}
n_{p}^{d}=n_{p}\cdot e^{-\left(\dfrac{h_{g}}{h_{c}}\right)^{2}},\quad p\in\Omega_{g}\ \label{dyn-on}
\end{equation}
\noindent with $h_{c}=0.02 \sqrt{2}$ \cite{piris-prl-2021}. The maximum value of $n_{p}^{d}$ is around 0.012 in accordance with the Pulay’s criterion that establishes an occupancy deviation of approximately 0.01 with respect to 1 or 0 for a natural orbital to contribute to the dynamic correlation. Clearly, GNOF does not take into account dynamic correlation of the single electrons ($p\in\Omega_{\mathrm{I}}$) via the $E_{dyn}^{inter}$ term. Considering real spatial orbitals ($L_{pq}=K_{pq}$) and $n_{p}\approx n_{p}^{d}$, it is not difficult to verify that the terms proportional to the product of the occupation numbers will cancel out, so that only those terms proportional to $\Pi$ will contribute significantly to the energy. 

It is important to note that GNOF preserves the total spin of the multiplet: $\mathrm{<}\hat{S}^{2}\mathrm{>}=S\left(S+1\right)$ \cite{Piris2019}. The solution is established by optimizing the energy with respect to the occupation numbers and to the natural orbitals, separately. Therefore, orbitals vary along the optimization process until the most favorable orbital interactions are found. All calculations have been carried out using the DoNOF code \cite{piris2021donof}. The STO-6G basis set \cite{sto-6g} was used throughout.

\section{Results}\label{sec:results}

\subsection{H$_{10}$ benchmark set}

\begin{figure}[ht]
\begin{centering}
\caption{\label{h10} Structure of the H$_{10}$ model systems studied in this work. \bigskip}
{\includegraphics[scale=0.38]{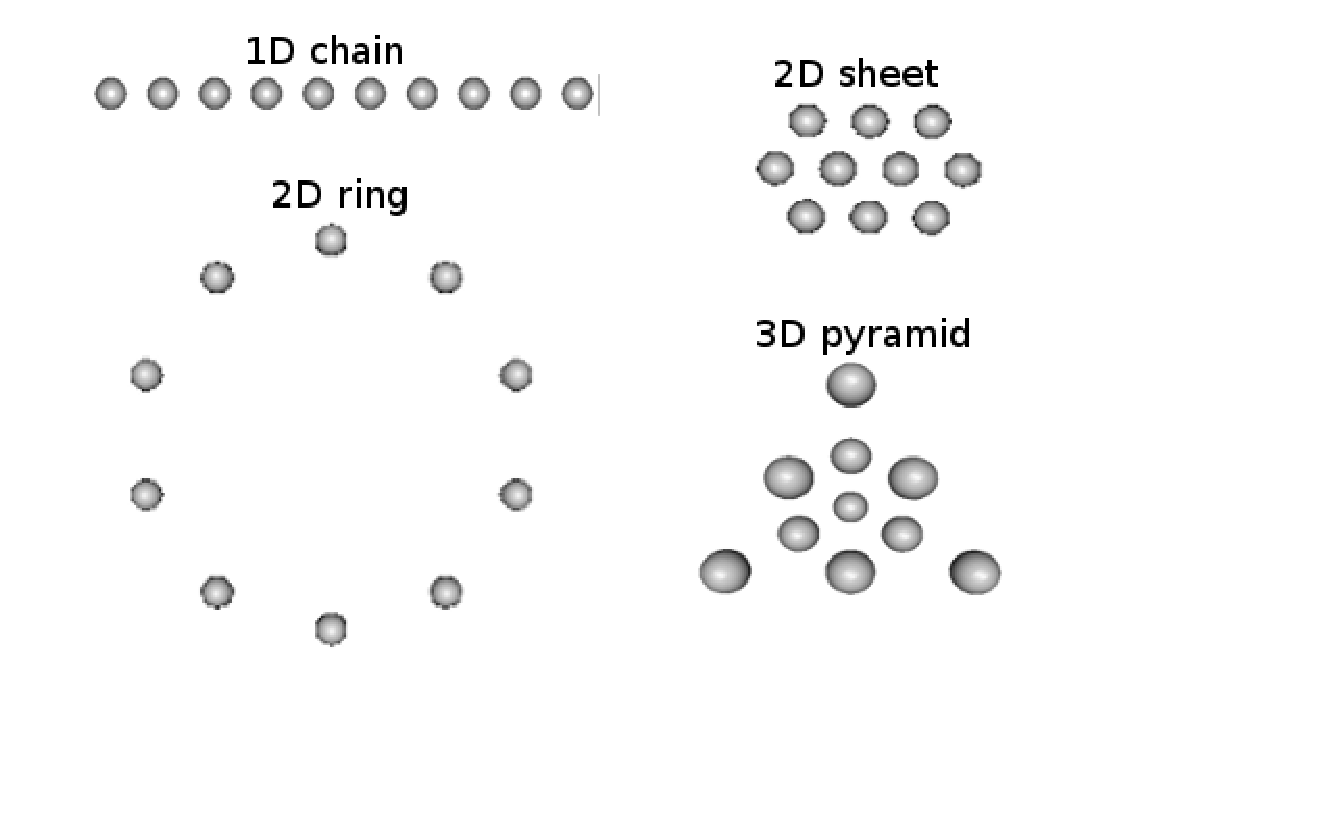}}
\end{centering}
\end{figure}

GNOF is analyzed in this section by using the H$_{10}$ benchmark set of 1D, 2D and 3D H clusters introduced by Stair and Evangelista \cite{stair-evangelista-jcp-2020} to test methods that converge to Full-Configuration Interaction (FCI) in the limit of no truncation of the wavefunction expansion. Fig. \ref{h10} shows the structure of the four H$_{10}$ model systems. The geometry of each model is controlled by the nearest neighbor H–H distance (in \r{A}). Recently \cite{li-eugene-deprince-jcp-2021}, Li, Liebenthal and DePrince employed this set in order to assess the effect of three-particle N-representability conditions (e.g., the T2 condition) not only on the energy curves, but also on a variety of correlation metrics including different order RDMs based quantities and the site-wise spin-spin correlation. Here, we compare reference FCI energies with PNOF7 and GNOF calculations. For comparison, in some examples we also report the v2RDM energies obtained in Ref. \cite{li-eugene-deprince-jcp-2021}.
It is worth noting that according to Ref. \cite{stair-evangelista-jcp-2020}, the reported correlation regimes correspond to \textit{U}/\textit{t} ranges from about $0.94$ at $r=0.75$ \r{A} to $8.55$ at $r=2.0$ \r{A}, where \textit{U} and \textit{t} are the repulsion and hopping parameters of the Hubbard Hamiltonian, respectively, and \textit{r} stands for the nearest neighbor H–H distance. PNOF7 has been largely proved to be an efficient method to study the Hubbard model in different filling situations, spin multiplicities, and many dimensions \cite{mitxelena2017performance,Mitxelena_2018-corrig,mitxelena2018a,mitxelena2020efficient1,mitxelena2020efficient2}. However, we have observed important energy differences with respect to FCI in the aforementioned \textit{U}/\textit{t} ranges due to the simultaneous presence of both weak and strong electron correlation effects. Hence, the here selected benchmark set constitutes a challenge for NOF approximations and is therefore suitable for testing whether GNOF is successful or not in dealing with weak and strong correlation in a more balanced way than its predecessors.

\begin{figure}[ht]
\begin{centering}
\caption{\label{Hchain} Dissociation energy curves corresponding to a 1D lineal chain of 10 H atoms. Energies (in Hartrees) obtained by using PNOF7, GNOF, FCI and v2RDM with two-particle DQG conditions. \bigskip\bigskip}
{\includegraphics[scale=0.35]{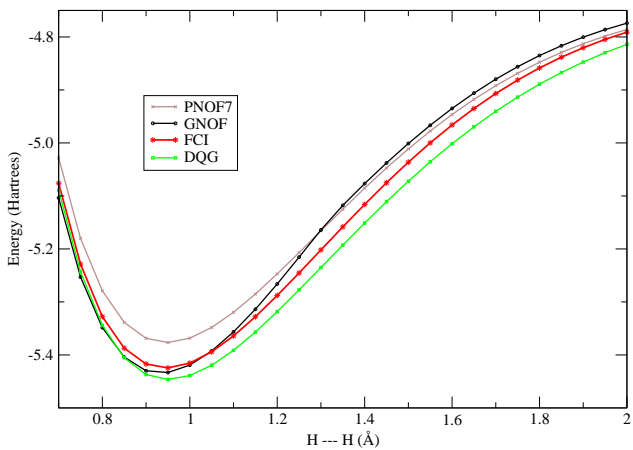}}
\end{centering}
\end{figure}

Fig. \ref{Hchain} shows the dissociation curves for a lineal chain of 10 H atoms. FCI and v2RDM with DQGT2 or three-particle complete positivity (3POS) conditions were found \cite{li-eugene-deprince-jcp-2021} to be in complete agreement, so we focus for this linear system on methods that consider only the two-particle N-representability conditions of the 2RDM. According to the results obtained for the 2D Hubbard model in Ref. \cite{mitxelena2020efficient2}, a NOF can outperform v2RDM with DQG conditions. However, although PNOF7 also outperformed v2RDM with three-particle T conditions for the 2D Hubbard Hamiltonian away from half-filling, this functional is less accurate in describing the dissociation of lineal H$_{10}$ in the equilibrium region. Noticeably, GNOF correctly recovers the lack of dynamical correlation shown by PNOF7 in this region. Although GNOF slightly underestimates the equilibrium distance, overall the dissociation curve and thereby the dissociation energy compares better with FCI than PNOF7 and its predecessors.

\begin{figure}[ht]
\begin{centering}
\caption{\label{Hsheet} Dissociation energy curves corresponding to a 2D sheet of 10 H atoms. Energies (in Hartrees) obtained by using FCI, v2RDM with two-particle DQG conditions (DQG), including T2 condition (DQGT2), and full three-particle positivity conditions (3POS). NOF approximations include PNOF7 and GNOF. \bigskip\bigskip\bigskip}
{\includegraphics[scale=0.35]{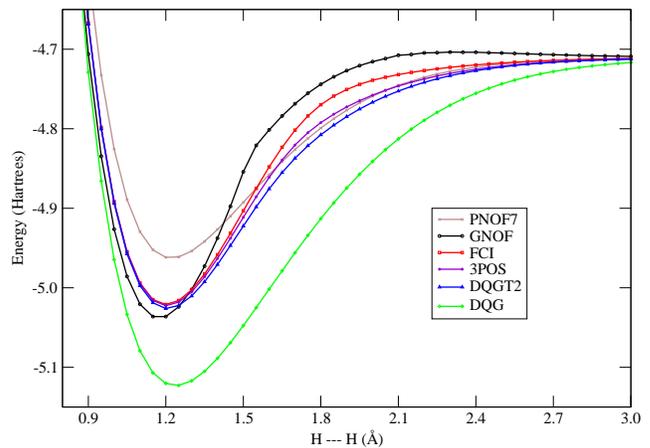}}
\end{centering}
\end{figure}

The addition of a spatial dimension increases the number of interatomic interactions, making the bond-breaking process more complex. 2D systems are known to be particularly challenging as the amount of electron correlation is higher than in their 1D counterparts. In Fig. \ref{Hsheet}, we show the dissociation curves for the 2D sheet of 10 H atoms shown in Fig. \ref{h10}. Since DQG yields too low energies along the whole dissociation curve, it is clear that the two-particle N-representability conditions are useful only if they are employed in an approximate 2RDM reconstruction with more restrictions, as occurs in the NOF approaches considered, namely PNOF7 and GNOF. PNOF7 shows good performance beyond interatomic distances of 1.5 \r{A}, where it embraces to the DQGT2 and 3POS curves. The latter accurately describe the equilibrium region but unfortunately produce energies well below FCI for distances greater than 1.5 \r{A}.

From Fig. \ref{Hsheet}, it can be seen that PNOF7 describes the dissociation of the 2D H sheet well, as it was observed for a 4x4 square grid of H atoms \cite{mitxelena2020efficient2}. Again, and similar to the result obtained for the linear array of Fig. \ref{Hchain}, the dynamic electron correlation present in the equilibrium region causes PNOF7 to stay very high compared to FCI energies. GNOF can recover the missing dynamic correlation in PNOF7, although it provides a slightly shorter equilibrium distance than FCI. It should be noted that GNOF remains close to FCI throughout the entire dissociation curve, showing a balanced treatment of dynamic and non-dynamic correlation effects.

\begin{figure}[ht]
\begin{centering}
\caption{\label{Hring} Dissociation energy curves corresponding to a 2D ring of 10 H atoms. Energies (in Hartrees) obtained by using PNOF7, GNOF, FCI and v2RDM with two-particle DQG conditions. \bigskip\bigskip\bigskip}
{\includegraphics[scale=0.35]{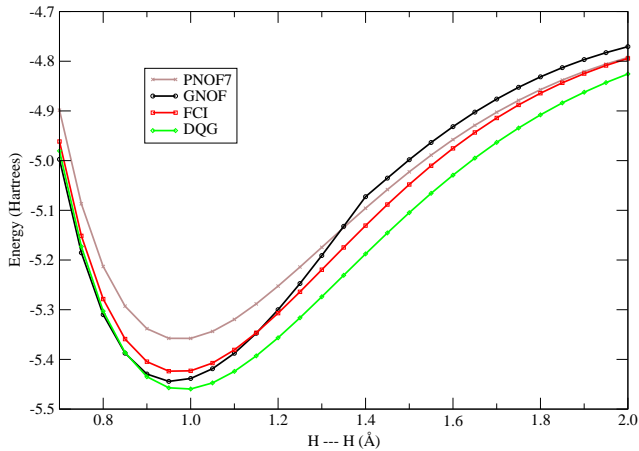}}
\end{centering}
\end{figure}

The dissociation of the 2D ring (see Fig. \ref{h10}) with respect to the nearest neighbor H–H distances shows similar characteristics of the dissociations discussed above. It is worth noting that we used this system to discuss the phase dilemma present in the development of PNOF7 \cite{mitxelena2018a}. First of all, the results obtained \cite{li-eugene-deprince-jcp-2021} with three-particle positivity conditions are very close to the FCI values, so they are not included in Fig. \ref{Hring}, as we did in the case of the linear chain of hydrogens. According to Fig. \ref{Hring}, PNOF7 remains above FCI throughout the curve, unlike what happens with DQG. GNOF recovers the dynamic correlation and improves the description of the equilibrium region with respect to known NOFs, as was the case with all systems studied so far.

In Figs. \ref{Hsheet} and \ref{Hring}, we must notice that black GNOF curves show a discontinuity of the derivative at 1.5 {\AA} and 1.4 {\AA}, respectively. It is well known that in the equilibrium region the dynamic correlation predominates and that in the dissociation asymptote the dominant character is given by the static correlation, while in the intermediate region both types of electron correlation compete. For both 2D systems, the GNOF potential energy curve (PEC) shows that the functional correctly describes the equilibrium and dissociation regions, but also suggests that in these regions we have two different solutions. These two different solutions have a non-smooth transition in the intermediate region, which is manifested in the discontinuity of the first derivative of the PEC with respect to the nearest neighbor H–H distances. In the case of the linear chain, both solutions also exist but GNOF is able to go from one solution to another smoothly. 

\begin{figure}[ht]
\begin{centering}
\caption{\label{homo} Occupancy ($2n_p$) of the lowest strongly occupied natural orbital.\bigskip\bigskip\bigskip}
\includegraphics[scale=0.33]{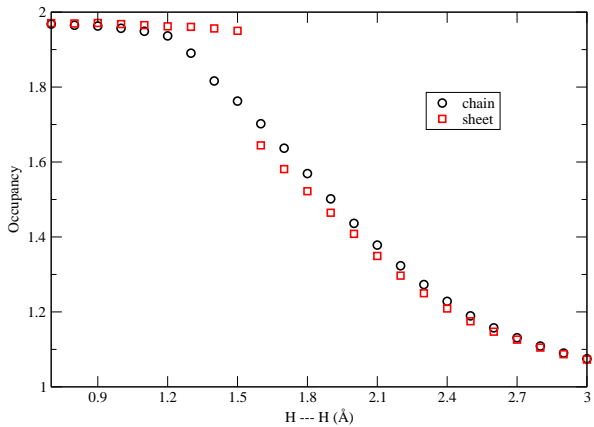}
\end{centering}
\end{figure}

In order to illustrate the transition between the two solutions, Fig. \ref{homo} shows the change in the occupancy corresponding to the strongly occupied natural orbital ($1.0 \le 2n_p \le 2.0$) with the lowest occupancy. Note that we report the occupancy for spatial orbitals and, hence, convergence to unity means that the orbitals are half-filled. In the case of the sheet, we can observe that the occupancy for the nearest neighbor H–H distances $\le 1.5$ {\AA} are close to 2, which reflects the dynamic character of these solutions in accordance with the Pulay's criterion. On the contrary, the solutions corresponding to greater distances have a marked non-dynamic character.

The existence of multiple solutions in the energy minimization problem for a NOF is expected given the non-linear character of the Euler equations \cite{piris2021donof}. When the correlation has a definite character one of the solutions is clearly of lower energy, however, when both types of correlation are important these solutions might be close. For the sheet, GNOF fails by not performing a correct balance of both types of correlation and that is why the discontinuity appears in the intermediate region. The proper PEC is obtained when there is a smooth transition between solutions, as observed in the case of the chain.

\begin{figure}[ht]
\begin{centering}
\caption{\label{Hpyramid} Dissociation energy curves corresponding to a 3D pyramid of 10 H atoms. Energies (in Hartrees) obtained by using FCI, v2RDM with two-particle DQG conditions (DQG), including T2 condition (DQGT2), and full three-particle positivity conditions (3POS). NOF approximations include PNOF7 and GNOF. \bigskip\bigskip\bigskip}
\includegraphics[scale=0.35]{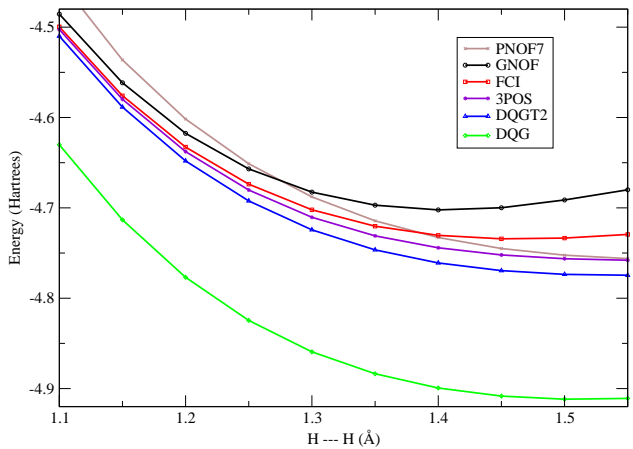}
\end{centering}
\end{figure}

The H$_{10}$ benchmark set is completed with a pyramid-shaped 3D cluster (see Fig. \ref{h10}). The energies obtained by using PNOF7, GNOF, FCI, and v2RDM with two- and three-particle positivity conditions, along the dissociation curves with respect to the nearest neighbor H–H distances, are shown in Fig. \ref{Hpyramid}. We can observe that three-particle constraints are not enough to avoid an overestimation of the correlation energy, and PNOF7 shows the same issue at interatomic distances greater than 1.4 \r{A}. All of these methods give a minimum well beyond FCI, which is located approximately at 1.46 \r{A}. Despite the relative error given by GNOF along the dissociation curve, its corresponding equilibrium distance of 1.41 \r{A} compares well with FCI, as well as the overall behavior along the entire curve. Thus, although a two-index approximation like GNOF does not recover all electron correlation, we can expect this functional to offer a more balanced treatment of the different types of electron correlation in complex three-dimensional physical and chemical problems.

\bigskip
\textit{Singlet-Triplet Gaps}\label{subsec:gaps}
\bigskip

In this subsection, we proceed to calculate single-triplet and singlet-quintet gaps corresponding to the H$_{10}$ benchmark set shown in Fig. \ref{h10} at nearest neighbor H–H distances of 1.0 \r{A} and 1.5 \r{A}. These were the distances used by Stair and Evangelista \cite{stair-evangelista-jcp-2020} to compute the density of states from the 50 lowest singlet, triplet, and quintet states of the H$_{10}$ systems. We focus on the gap calculated only from the minimum energy states corresponding to each spin. Recall that we use the recently introduced \cite{Piris2019} formalism for spin multiplets in the framework of NOF theory that guarantees the preservation of the total spin of the system, in addition to its projection.

\begin{table}[h!]
  \begin{center}
    \caption{Lowest energy singlet-triplet and singlet-quintet gaps (in Hartrees) at 
    nearest neighbor H–H distance of 1.0 \r{A}. The FCI values are inferred from Fig. 2 of Ref. \cite{stair-evangelista-jcp-2020} \bigskip}
    \label{tab:table-gaps-1}
    \begin{tabular}{l | c c | c r}
\multirow{2}{*} & \multicolumn{2}{c|}{S - T} & \multicolumn{2}{c}{ S - Q}  \\ 
      system & FCI & GNOF & FCI & GNOF \\
      \hline
      chain   & 0.12 & 0.13 & 0.47 & 0.49 \\
      ring    & 0.24 & 0.29 &  -   & 0.58 \\
      sheet   & 0.22 & 0.24 &  -   & 0.61 \\
      pyramid & 0.04 & 0.01 &  -   & 0.56 \\
    \end{tabular}
  \end{center}
\end{table}

Tables \ref{tab:table-gaps-1} and \ref{tab:table-gaps-15} show the lowest energy singlet-triplet and singlet-quintet gaps obtained using GNOF, and the FCI values inferred from Fig. 2 of Ref. \cite{stair-evangelista-jcp-2020}. At shorter bond length, the 1D and 2D systems show large gaps between the ground state and the lowest triplet and quintet states, whereas at longer bond length, the singlet–triplet and singlet-quintet gaps decrease for all systems. Interestingly, the 3D pyramid shows a small singlet-triplet gap at 1.0 \r{A} and an almost zero gap at 1.5 \r{A}. In fact, GNOF predicts that the lowest triplet state is negligibly lower than the singlet state. We observe a reasonable agreement between the gaps obtained with both methods, especially at 1.5 \r{A}.

\begin{table}[h!]
  \begin{center}
    \caption{Lowest energy singlet-triplet and singlet-quintet gaps (in Hartrees) at 
    nearest neighbor H–H distance of 1.5 \r{A}. The FCI values are inferred from Fig. 2 of Ref. \cite{stair-evangelista-jcp-2020} \bigskip}
    \label{tab:table-gaps-15}
    \begin{tabular}{l | c c | c r}
\multirow{2}{*} & \multicolumn{2}{c|}{S - T} & \multicolumn{2}{c}{ S - Q}  \\ 
      system         & FCI & GNOF & FCI & GNOF \\
      \hline
      chain   & 0.03  &  0.05  & 0.14 & 0.13 \\
      ring    & 0.05  &  0.06  & 0.17 & 0.16 \\
      sheet   & 0.06  &  0.06  & 0.15 & 0.13 \\
      pyramid & 0.005 & -0.001 & 0.16 & 0.15 \\
    \end{tabular}
  \end{center}
\end{table}

\bigskip

\subsection{Large hydrogen clusters}\label{subsec:large}

In this section, we discuss our results on large hydrogen clusters, namely a 1D H$_{50}$ chain and a 4x4x4 3D hydrogen cube. The size of these systems makes FCI calculations not possible, so density-matrix renormalization group (DMRG) method is used as reference whenever it is available \cite{hachman-dmrg-2006}. Hydrogen clusters exhibit electron localization when their atoms are symmetrically dissociated. Thus, we stretch all internuclear distances between adjacent atoms simultaneously to obtain, at the end, 50 and 64 isolated hydrogen atoms, respectively for the 1D and 3D systems. Similar to the Hubbard model when the correlation regime is increased, a metal-to-insulator transition is observed in both systems. Accordingly, they are paradigmatic models for strongly correlated Mott insulators.

\begin{figure}[ht]
\begin{centering}
\caption{\label{H50} Symmetric dissociation energy curves corresponding to a 1D H$_{50}$ chain. Energies per atom (in Hartrees) obtained by using PNOF7, GNOF, DMRG and v2RDM with two-particle DQG conditions (DQG) \bigskip\bigskip\bigskip}
\includegraphics[scale=0.35]{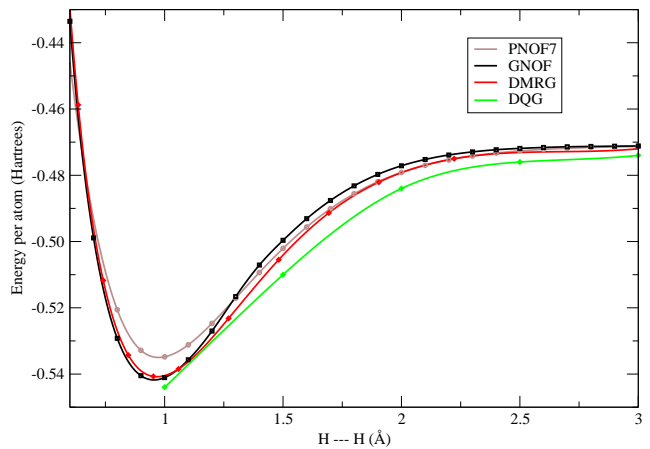}
\end{centering}
\end{figure}

Symmetric dissociation energy curves corresponding to PNOF7, GNOF, DMRG and v2RDM with DQG constraints are shown in Fig. \ref{H50} for the H$_{50}$ chain. Note that DQG values were inferred from Fig. 2 of Ref. \cite{mazziotti2011H50}. The H$_{50}$ was studied in detail in Ref. \cite{mitxelena2020efficient1} in the context of PNOF7. Despite it compares well with reference DMRG energies along the dissociation curve and it exhibits a correct behaviour of natural occupation numbers (see Fig. 3 from Ref. \cite{mitxelena2020efficient1}), PNOF7 lacks dynamical correlation at the equilibrium region. As already observed in the H$_{10}$ benchmark set, GNOF not only improves the dissociation energy given by PNOF7, but it also sticks to DMRG along the full curve. More importantly, GNOF accuracy and its ability to retrieve the dynamic correlation missing in PNOF7 do not deteriorate going from small to large clusters. Electron correlation is thus correctly accounted by considering the dynamic part of the occupation numbers in Eqs. \ref{edyn} and \ref{dyn-on}. 

\begin{figure}[ht]
\begin{centering}
\caption{\label{H64} Symmetric dissociation energy curves corresponding to a 4x4x4 hydrogen cube. Energies per atom (in Hartrees) obtained by using RHF, PNOF7, MP2, GNOF and v2RDM with two-particle DQG conditions (DQG). \bigskip\bigskip\bigskip}
\includegraphics[scale=0.33]{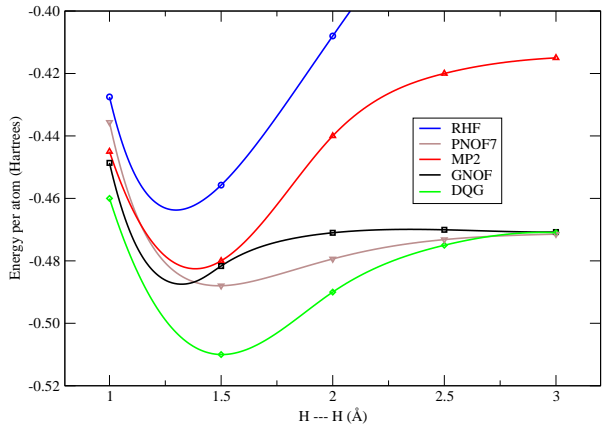}
\end{centering}
\end{figure}

Regarding the performance of GNOF for a large cluster with multiple dimensions, Fig. \ref{H64} shows the symmetric dissociation energy curves corresponding to RHF, MP2, PNOF7, GNOF and v2RDM with DQG constraints for a 4x4x4 hydrogen cube. Here, the MP2 and DQG values were inferred from Fig. 6 of Ref. \cite{mazziotti2010}.  Similar to the result obtained for the 3D pyramid in Fig. \ref{Hpyramid}, both PNOF7 and DQG overestimate the equilibrium distance. It should be noted that the dissociation energy of PNOF7 is similar to that of GNOF, but at the cost of finding a solution with a markedly non-dynamic character indicated by the occupation numbers obtained, which is unexpected for the equilibrium region. GNOF, in contrast, provides an equilibrium bond distance between RHF and MP2, and dissociates to the PNOF7 limit also together with DQG. Unfortunately, we do not have a reliable reference curve to adequately test the accuracy of the method.

To end this section we come up with a study of the metal-to-insulator transition observed in large systems. This transition has been previously studied \cite{mitxelena2017performance,mitxelena2020efficient2} in the framework of PNOF methods for the Hubbard model, either looking at the natural occupation numbers or studying the 1RDM in the site-basis. In the case of hydrogen clusters, it is convenient to compute the average of off-diagonal 1RDM elements. We employ the harmonic average $\gamma$ of all the off-diagonal 1RDM elements $\Gamma ^{AO}_{ij}$ in the atomic orbital basis set \cite{mazziotti2010} to describe quantitatively the metal-insulator transition, namely

\begin{equation}
\gamma = \sqrt {\dfrac{1}{\mathrm{N}\left(\mathrm{N}-1\right)} \sum\limits _{i\neq j} (\Gamma ^{AO}_{ij})^{2} }\label{eq:harmonic-average}
\end{equation}

\begin{figure}[ht]
\begin{centering}
\caption{\label{gamma} Metal-to-insulator transition in the 4x4x4 hydrogen cube under the change of the nearest neighbor H–H distance. \bigskip\bigskip\bigskip}
\includegraphics[scale=0.3]{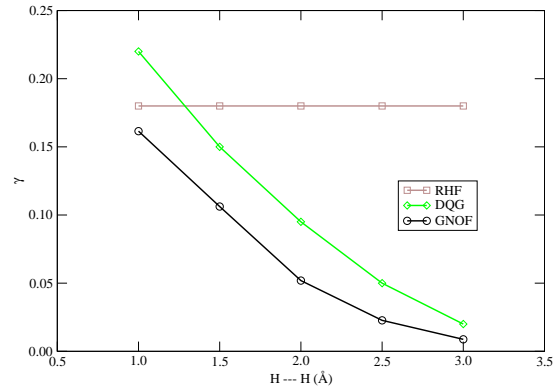}
\end{centering}
\end{figure}

Fig. \ref{gamma} presents the dependence of $\gamma$ on the H–H distances for the 4x4x4 cube. RHF and DQG values were obtained from Fig. 7 of Ref. \cite{mazziotti2010}. The loss of spatial correlation with increasing separation causes the off-diagonal $\Gamma^{AO}_{ij}$ to approach zero, and thus the harmonic average $\gamma$ also be close to zero. GNOF goes quantitatively parallel to DQG, as a lower bound to the latter. Similar behavior was obtained in Ref. \cite{mazziotti2010} for the FCI with respect to DQG for the H$_{14}$ chain. Unfortunately, the lack of a reference does not allow us to be conclusive about the numerical accuracy of the curves shown in this figure. Anyway, the metal-to-insulator transition is well captured by GNOF for large 3D systems.

\section{Conclusions}\label{sec:conclusions}

In this work, we have tested our recently proposed global natural orbital functional (GNOF) against the full configuration interaction (FCI) and density-matrix renormalization group (DMRG) methods by using a series of benchmark hydrogen models with a tunable degree of correlation. Although in some regions GNOF shows larger relative errors than variational 2RDM calculations with three-particle N-representability constraints, this functional has shown correct performance along the complete dissociation curves for all studied systems in one, two and three dimensions. The GNOF curves remain close to those of FCI and DMRG, so its treatment of dynamic and static (non-dynamic or strong) electronic correlations is more balanced than in other methods studied throughout the article. But there is still room for improvement. In the 2D systems we observed a discontinuity of the first derivative in the dissociation curves at intermediate distances. In the future, we must achieve for these systems a smooth transition between the solutions that characterize a specific type of correlation since in the intermediate region both types of correlation are equally important.

Previous attempts to recover full electron correlation within the framework of first-order reduced density matrix (1RDM) theory rely on hybrid approaches, such as combining NOF with perturbation theories \cite{Piris2017,piris-2018-dyn,hollet-titou-2020,mauricio-2021-jctc} or with the DFT \cite{wenna-jpcl-2022,elayan-hollet-2022}. Unfortunately, hybrid approaches only give access to the energy. It is preferable to develop a functional that includes from the outset all the electronic correlation, so that we also have access to the correlated reduced density matrices and orbitals. The present work represents a great step forward to establish NOF among electronic structure methods for global electronic correlation problems where mean-field approaches are not accurate enough and it is necessary to resort to multiple reference methods to correctly handle both types of correlation. Recall that these latter techniques are usually computationally expensive and require prior knowledge of the system. NOF methods, on the contrary, have shown \cite{piris2021donof} a favorable computational scaling of M to the fourth power times a prefactor, M being the number of basis functions. In the case of GNOF, this prefactor corresponds to the number of strongly occupied orbitals, i.e. $\mathrm{\mathrm{N}_{\Omega}=}\mathrm{N_{II}}/2+\mathrm{N_{I}}$ (see section \ref{sec:theory} above). Recently \cite{lee-yee-piris-RI}, the resolution of the identity (RI) approximation has been implemented in DoNOF, reporting a quantitative study of the computational time required to carry out NOF calculations in increasing size real molecules. The DoNOF-RI implementation reduces the memory and arithmetic scaling factors in NOF calculations.

The reported values of the singlet-triplet and singlet-quintet gaps demonstrate the possibility of performing spin multiplet calculations conserving the total spin of the system with a NOF. Our formalism differs from the methods commonly used in electronic structure calculations that focus on the high-spin component or break the spin symmetry of the system. 

It has also been shown that GNOF can describe the metal-to-insulator transition in large 1D and 3D hydrogen systems. GNOF symmetric dissociation energy curves, as well as the dependence of the harmonic average of all the off-diagonal 1RDM elements on the bond stretching, exhibit the expected behaviour for both the H$_{50}$ chain and the 4x4x4 H$_{64}$ cube.

The benchmark set used in this work involves only minimal basis that allowed us to make comparisons with accurate FCI and DMRG calculations. Our previous calculations with larger basis sets, e.g. triple zeta basis sets, showed excellent performance of GNOF, so we expect the conclusions obtained here to hold for strongly correlated systems in studies with larger basis with better recovery of the dynamic electron correlation. 
\begin{acknowledgments}
The authors thank for technical and human support provided by IZO-SGI SGIker of UPV/EHU and European funding (ERDF and ESF) and DIPC for the generous allocation of computational resources. Financial support comes from the Eusko Jaurlaritza (Basque Government), Ref.: IT1254-19.
\end{acknowledgments}


\begin{thebibliography}{42}
\expandafter\ifx\csname natexlab\endcsname\relax\def\natexlab#1{#1}\fi
\expandafter\ifx\csname bibnamefont\endcsname\relax
  \def\bibnamefont#1{#1}\fi
\expandafter\ifx\csname bibfnamefont\endcsname\relax
  \def\bibfnamefont#1{#1}\fi
\expandafter\ifx\csname citenamefont\endcsname\relax
  \def\citenamefont#1{#1}\fi
\expandafter\ifx\csname url\endcsname\relax
  \def\url#1{\texttt{#1}}\fi
\expandafter\ifx\csname urlprefix\endcsname\relax\def\urlprefix{URL }\fi
\providecommand{\bibinfo}[2]{#2}
\providecommand{\eprint}[2][]{\url{#2}}

\bibitem[{\citenamefont{Tew et~al.}(2007)\citenamefont{Tew, Klopper, and
  Helgaker}}]{tew-jcc-2007}
\bibinfo{author}{\bibfnamefont{D.~P.} \bibnamefont{Tew}},
  \bibinfo{author}{\bibfnamefont{W.}~\bibnamefont{Klopper}}, \bibnamefont{and}
  \bibinfo{author}{\bibfnamefont{T.}~\bibnamefont{Helgaker}},
  \bibinfo{journal}{J. Comput. Chem.} \textbf{\bibinfo{volume}{28}},
  \bibinfo{pages}{1307} (\bibinfo{year}{2007}).

\bibitem[{\citenamefont{Cohen et~al.}(2012)\citenamefont{Cohen, Mori-Sánchez,
  and Yang}}]{cohen-dft-2012}
\bibinfo{author}{\bibfnamefont{A.~J.} \bibnamefont{Cohen}},
  \bibinfo{author}{\bibfnamefont{P.}~\bibnamefont{Mori-Sánchez}},
  \bibnamefont{and} \bibinfo{author}{\bibfnamefont{W.}~\bibnamefont{Yang}},
  \bibinfo{journal}{Chem. Rev.} \textbf{\bibinfo{volume}{112}},
  \bibinfo{pages}{289} (\bibinfo{year}{2012}).

\bibitem[{\citenamefont{Ai et~al.}(2022)\citenamefont{Ai, Fang, and
  Su}}]{wenna-jpcl-2022}
\bibinfo{author}{\bibfnamefont{W.}~\bibnamefont{Ai}},
  \bibinfo{author}{\bibfnamefont{W.-H.} \bibnamefont{Fang}}, \bibnamefont{and}
  \bibinfo{author}{\bibfnamefont{N.~Q.} \bibnamefont{Su}}, \bibinfo{journal}{J.
  Phys. Chem. Lett.} \textbf{\bibinfo{volume}{13}}, \bibinfo{pages}{1744}
  (\bibinfo{year}{2022}).

\bibitem[{\citenamefont{Kirkpatrick et~al.}(2021)\citenamefont{Kirkpatrick,
  McMorrow, Turban, Gaunt, Spencer, Matthews, Obika, Thiry, Fortunato, Pfau
  et~al.}}]{science-2021-frac-dft}
\bibinfo{author}{\bibfnamefont{J.}~\bibnamefont{Kirkpatrick}},
  \bibinfo{author}{\bibfnamefont{B.}~\bibnamefont{McMorrow}},
  \bibinfo{author}{\bibfnamefont{D.~H.~P.} \bibnamefont{Turban}},
  \bibinfo{author}{\bibfnamefont{A.~L.} \bibnamefont{Gaunt}},
  \bibinfo{author}{\bibfnamefont{J.~S.} \bibnamefont{Spencer}},
  \bibinfo{author}{\bibfnamefont{A.~G. D.~G.} \bibnamefont{Matthews}},
  \bibinfo{author}{\bibfnamefont{A.}~\bibnamefont{Obika}},
  \bibinfo{author}{\bibfnamefont{L.}~\bibnamefont{Thiry}},
  \bibinfo{author}{\bibfnamefont{M.}~\bibnamefont{Fortunato}},
  \bibinfo{author}{\bibfnamefont{D.}~\bibnamefont{Pfau}}, \bibnamefont{et~al.},
  \bibinfo{journal}{Science} \textbf{\bibinfo{volume}{374}},
  \bibinfo{pages}{1385} (\bibinfo{year}{2021}).

\bibitem[{\citenamefont{Mazziotti}(2012{\natexlab{a}})}]{mazziotti2012review}
\bibinfo{author}{\bibfnamefont{D.~A.} \bibnamefont{Mazziotti}},
  \bibinfo{journal}{Chem. Rev.} \textbf{\bibinfo{volume}{112}},
  \bibinfo{pages}{244–262} (\bibinfo{year}{2012}{\natexlab{a}}).

\bibitem[{\citenamefont{Fosso-Tande et~al.}(2016)\citenamefont{Fosso-Tande,
  Nguyen, Gidofalvi, and DePrince}}]{FossoTande2016}
\bibinfo{author}{\bibfnamefont{J.}~\bibnamefont{Fosso-Tande}},
  \bibinfo{author}{\bibfnamefont{T.-S.} \bibnamefont{Nguyen}},
  \bibinfo{author}{\bibfnamefont{G.}~\bibnamefont{Gidofalvi}},
  \bibnamefont{and} \bibinfo{author}{\bibfnamefont{A.~E.}
  \bibnamefont{DePrince}}, \bibinfo{journal}{J. Chem. Theory Comput.}
  \textbf{\bibinfo{volume}{12}}, \bibinfo{pages}{2260} (\bibinfo{year}{2016}).

\bibitem[{\citenamefont{Piris}(2017)}]{Piris2017}
\bibinfo{author}{\bibfnamefont{M.}~\bibnamefont{Piris}},
  \bibinfo{journal}{Phys. Rev. Lett.} \textbf{\bibinfo{volume}{119}},
  \bibinfo{pages}{063002} (\bibinfo{year}{2017}).

\bibitem[{\citenamefont{Schouten et~al.}(2021)\citenamefont{Schouten, Sager, ,
  and Mazziotti}}]{mazziotti2021}
\bibinfo{author}{\bibfnamefont{A.~O.} \bibnamefont{Schouten}},
  \bibinfo{author}{\bibfnamefont{L.~M.} \bibnamefont{Sager}}, ,
  \bibnamefont{and} \bibinfo{author}{\bibfnamefont{D.~A.}
  \bibnamefont{Mazziotti}}, \bibinfo{journal}{J. Phys. Chem. Lett.}
  \textbf{\bibinfo{volume}{12}}, \bibinfo{pages}{9906–9911}
  (\bibinfo{year}{2021}).

\bibitem[{\citenamefont{Piris}(2021)}]{piris-prl-2021}
\bibinfo{author}{\bibfnamefont{M.}~\bibnamefont{Piris}},
  \bibinfo{journal}{Phys. Rev. Lett.} \textbf{\bibinfo{volume}{127}},
  \bibinfo{pages}{233001} (\bibinfo{year}{2021}).

\bibitem[{\citenamefont{Mazziotti}(2012{\natexlab{b}})}]{mazziotti2012structure}
\bibinfo{author}{\bibfnamefont{D.~A.} \bibnamefont{Mazziotti}},
  \bibinfo{journal}{Phys. Rev. Lett.} \textbf{\bibinfo{volume}{108}},
  \bibinfo{pages}{263002} (\bibinfo{year}{2012}{\natexlab{b}}).

\bibitem[{\citenamefont{Mazziotti}(2006)}]{mazziotti2006}
\bibinfo{author}{\bibfnamefont{D.~A.} \bibnamefont{Mazziotti}},
  \bibinfo{journal}{Phys. Rev. A} \textbf{\bibinfo{volume}{74}},
  \bibinfo{pages}{032501} (\bibinfo{year}{2006}).

\bibitem[{\citenamefont{Mazziotti}(2007)}]{mazziotti2007}
\bibinfo{author}{\bibfnamefont{D.~A.} \bibnamefont{Mazziotti}},
  \bibinfo{journal}{J. Chem. Phys.} \textbf{\bibinfo{volume}{126}},
  \bibinfo{pages}{024105} (\bibinfo{year}{2007}).

\bibitem[{\citenamefont{Verstichel et~al.}(2012)\citenamefont{Verstichel, van
  Aggelen, Poelmans, and Van~Neck}}]{vertichel-prl-2012}
\bibinfo{author}{\bibfnamefont{B.}~\bibnamefont{Verstichel}},
  \bibinfo{author}{\bibfnamefont{H.}~\bibnamefont{van Aggelen}},
  \bibinfo{author}{\bibfnamefont{W.}~\bibnamefont{Poelmans}}, \bibnamefont{and}
  \bibinfo{author}{\bibfnamefont{D.}~\bibnamefont{Van~Neck}},
  \bibinfo{journal}{Phys. Rev. Lett.} \textbf{\bibinfo{volume}{108}},
  \bibinfo{pages}{213001} (\bibinfo{year}{2012}).

\bibitem[{\citenamefont{Rubin and Mazziotti}(2014)}]{rubin-mazziotti-tca-2014}
\bibinfo{author}{\bibfnamefont{N.~H.} \bibnamefont{Rubin}} \bibnamefont{and}
  \bibinfo{author}{\bibfnamefont{D.~A.} \bibnamefont{Mazziotti}},
  \bibinfo{journal}{Theor. Chem. Acc.} \textbf{\bibinfo{volume}{133}},
  \bibinfo{pages}{1492} (\bibinfo{year}{2014}).

\bibitem[{\citenamefont{Li et~al.}(2021)\citenamefont{Li, Liebenthal, and
  DePrince}}]{li-eugene-deprince-jcp-2021}
\bibinfo{author}{\bibfnamefont{R.~R.} \bibnamefont{Li}},
  \bibinfo{author}{\bibfnamefont{M.~D.} \bibnamefont{Liebenthal}},
  \bibnamefont{and} \bibinfo{author}{\bibfnamefont{A.~E.}
  \bibnamefont{DePrince}}, \bibinfo{journal}{J. Chem. Phys.}
  \textbf{\bibinfo{volume}{155}}, \bibinfo{pages}{174110}
  (\bibinfo{year}{2021}).

\bibitem[{\citenamefont{Mazziotti}(2016)}]{mazziotti2016}
\bibinfo{author}{\bibfnamefont{D.~A.} \bibnamefont{Mazziotti}},
  \bibinfo{journal}{Phys. Rev. Lett.} \textbf{\bibinfo{volume}{117}},
  \bibinfo{pages}{153001} (\bibinfo{year}{2016}).

\bibitem[{\citenamefont{Mazziotti}(2020)}]{mazziotti2020}
\bibinfo{author}{\bibfnamefont{D.~A.} \bibnamefont{Mazziotti}},
  \bibinfo{journal}{Phys. Rev. A} \textbf{\bibinfo{volume}{102}},
  \bibinfo{pages}{052819} (\bibinfo{year}{2020}).

\bibitem[{\citenamefont{Mitxelena and
  Piris}(2020{\natexlab{a}})}]{mitxelena2020efficient1}
\bibinfo{author}{\bibfnamefont{I.}~\bibnamefont{Mitxelena}} \bibnamefont{and}
  \bibinfo{author}{\bibfnamefont{M.}~\bibnamefont{Piris}}, \bibinfo{journal}{J.
  Phys. Condens. Matter} \textbf{\bibinfo{volume}{32}}, \bibinfo{pages}{17LT01}
  (\bibinfo{year}{2020}{\natexlab{a}}).

\bibitem[{\citenamefont{Mitxelena and
  Piris}(2020{\natexlab{b}})}]{mitxelena2020efficient2}
\bibinfo{author}{\bibfnamefont{I.}~\bibnamefont{Mitxelena}} \bibnamefont{and}
  \bibinfo{author}{\bibfnamefont{M.}~\bibnamefont{Piris}}, \bibinfo{journal}{J.
  Chem. Phys.} \textbf{\bibinfo{volume}{152}}, \bibinfo{pages}{064108}
  (\bibinfo{year}{2020}{\natexlab{b}}).

\bibitem[{\citenamefont{Piris}(2006)}]{Piris2006}
\bibinfo{author}{\bibfnamefont{M.}~\bibnamefont{Piris}}, \bibinfo{journal}{Int.
  J. Quantum Chem.} \textbf{\bibinfo{volume}{106}}, \bibinfo{pages}{1093}
  (\bibinfo{year}{2006}).

\bibitem[{\citenamefont{Piris et~al.}(2011)\citenamefont{Piris, Lopez,
  Ruip{\'e}rez, Matxain, and Ugalde}}]{piris2011-pnof5}
\bibinfo{author}{\bibfnamefont{M.}~\bibnamefont{Piris}},
  \bibinfo{author}{\bibfnamefont{X.}~\bibnamefont{Lopez}},
  \bibinfo{author}{\bibfnamefont{F.}~\bibnamefont{Ruip{\'e}rez}},
  \bibinfo{author}{\bibfnamefont{J.}~\bibnamefont{Matxain}}, \bibnamefont{and}
  \bibinfo{author}{\bibfnamefont{J.}~\bibnamefont{Ugalde}},
  \bibinfo{journal}{J. Chem. Phys.} \textbf{\bibinfo{volume}{134}},
  \bibinfo{pages}{164102} (\bibinfo{year}{2011}).

\bibitem[{\citenamefont{Piris et~al.}(2013)\citenamefont{Piris, Matxain, and
  Lopez}}]{Piris2013e}
\bibinfo{author}{\bibfnamefont{M.}~\bibnamefont{Piris}},
  \bibinfo{author}{\bibfnamefont{J.~M.} \bibnamefont{Matxain}},
  \bibnamefont{and} \bibinfo{author}{\bibfnamefont{X.}~\bibnamefont{Lopez}},
  \bibinfo{journal}{J. Chem. Phys.} \textbf{\bibinfo{volume}{139}},
  \bibinfo{pages}{234109} (\bibinfo{year}{2013}).

\bibitem[{\citenamefont{L{\"o}wdin and Shull}(1956)}]{lowdin:56pr}
\bibinfo{author}{\bibfnamefont{P.-O.} \bibnamefont{L{\"o}wdin}}
  \bibnamefont{and} \bibinfo{author}{\bibfnamefont{H.}~\bibnamefont{Shull}},
  \bibinfo{journal}{Phys. Rev.} \textbf{\bibinfo{volume}{101}},
  \bibinfo{pages}{1730} (\bibinfo{year}{1956}).

\bibitem[{\citenamefont{Pernal}(2013)}]{Pernal2013}
\bibinfo{author}{\bibfnamefont{K.}~\bibnamefont{Pernal}},
  \bibinfo{journal}{Comp. Theor. Chem.} \textbf{\bibinfo{volume}{1003}},
  \bibinfo{pages}{127} (\bibinfo{year}{2013}).

\bibitem[{\citenamefont{Matxain et~al.}(2011)\citenamefont{Matxain, Piris,
  Ruip{\'e}rez, Lopez, and Ugalde}}]{matxain2011-pnof5}
\bibinfo{author}{\bibfnamefont{J.}~\bibnamefont{Matxain}},
  \bibinfo{author}{\bibfnamefont{M.}~\bibnamefont{Piris}},
  \bibinfo{author}{\bibfnamefont{F.}~\bibnamefont{Ruip{\'e}rez}},
  \bibinfo{author}{\bibfnamefont{X.}~\bibnamefont{Lopez}}, \bibnamefont{and}
  \bibinfo{author}{\bibfnamefont{J.}~\bibnamefont{Ugalde}},
  \bibinfo{journal}{Phys. Chem. Chem. Phys.} \textbf{\bibinfo{volume}{13}},
  \bibinfo{pages}{20129} (\bibinfo{year}{2011}).

\bibitem[{\citenamefont{Mitxelena et~al.}(2017)\citenamefont{Mitxelena, Piris,
  and Rodr{\'\i}guez-Mayorga}}]{mitxelena2017performance}
\bibinfo{author}{\bibfnamefont{I.}~\bibnamefont{Mitxelena}},
  \bibinfo{author}{\bibfnamefont{M.}~\bibnamefont{Piris}}, \bibnamefont{and}
  \bibinfo{author}{\bibfnamefont{M.}~\bibnamefont{Rodr{\'\i}guez-Mayorga}},
  \bibinfo{journal}{J. Phys. Condens. Matter} \textbf{\bibinfo{volume}{29}},
  \bibinfo{pages}{425602} (\bibinfo{year}{2017}).

\bibitem[{\citenamefont{Mitxelena
  et~al.}(2018{\natexlab{a}})\citenamefont{Mitxelena, Rodr{\'{i}}guez-Mayorga,
  and Piris}}]{mitxelena2018a}
\bibinfo{author}{\bibfnamefont{I.}~\bibnamefont{Mitxelena}},
  \bibinfo{author}{\bibfnamefont{M.}~\bibnamefont{Rodr{\'{i}}guez-Mayorga}},
  \bibnamefont{and} \bibinfo{author}{\bibfnamefont{M.}~\bibnamefont{Piris}},
  \bibinfo{journal}{Eur. Phys. J. B} \textbf{\bibinfo{volume}{91}},
  \bibinfo{pages}{109} (\bibinfo{year}{2018}{\natexlab{a}}).

\bibitem[{\citenamefont{Bardeen et~al.}(1957)\citenamefont{Bardeen, Cooper, and
  Schrieffer}}]{Superconductivity1}
\bibinfo{author}{\bibfnamefont{J.}~\bibnamefont{Bardeen}},
  \bibinfo{author}{\bibfnamefont{L.~N.} \bibnamefont{Cooper}},
  \bibnamefont{and} \bibinfo{author}{\bibfnamefont{J.~R.}
  \bibnamefont{Schrieffer}}, \bibinfo{journal}{Phys. Rev.}
  \textbf{\bibinfo{volume}{108}}, \bibinfo{pages}{1175} (\bibinfo{year}{1957}).

\bibitem[{\citenamefont{Mitxelena
  et~al.}(2018{\natexlab{b}})\citenamefont{Mitxelena, Piris, and
  Rodr{\'{\i}}guez-Mayorga}}]{Mitxelena_2018-corrig}
\bibinfo{author}{\bibfnamefont{I.}~\bibnamefont{Mitxelena}},
  \bibinfo{author}{\bibfnamefont{M.}~\bibnamefont{Piris}}, \bibnamefont{and}
  \bibinfo{author}{\bibfnamefont{M.}~\bibnamefont{Rodr{\'{\i}}guez-Mayorga}},
  \bibinfo{journal}{J. Phys.: Condens. Matter} \textbf{\bibinfo{volume}{30}},
  \bibinfo{pages}{089501} (\bibinfo{year}{2018}{\natexlab{b}}).

\bibitem[{\citenamefont{Sinitskiy et~al.}(2010)\citenamefont{Sinitskiy,
  Greenman, and Mazziotti}}]{mazziotti2010}
\bibinfo{author}{\bibfnamefont{A.~V.} \bibnamefont{Sinitskiy}},
  \bibinfo{author}{\bibfnamefont{L.}~\bibnamefont{Greenman}}, \bibnamefont{and}
  \bibinfo{author}{\bibfnamefont{D.~A.} \bibnamefont{Mazziotti}},
  \bibinfo{journal}{J. Chem. Phys.} \textbf{\bibinfo{volume}{133}},
  \bibinfo{pages}{014104} (\bibinfo{year}{2010}).

\bibitem[{\citenamefont{Motta et~al.}(2017)\citenamefont{Motta, Ceperley, Chan,
  Gomez, Gull, Guo, Jim\'enez-Hoyos, Lan, Li, Ma et~al.}}]{motta-prx-2017}
\bibinfo{author}{\bibfnamefont{M.}~\bibnamefont{Motta}},
  \bibinfo{author}{\bibfnamefont{D.~M.} \bibnamefont{Ceperley}},
  \bibinfo{author}{\bibfnamefont{G.~K.-L.} \bibnamefont{Chan}},
  \bibinfo{author}{\bibfnamefont{J.~A.} \bibnamefont{Gomez}},
  \bibinfo{author}{\bibfnamefont{E.}~\bibnamefont{Gull}},
  \bibinfo{author}{\bibfnamefont{S.}~\bibnamefont{Guo}},
  \bibinfo{author}{\bibfnamefont{C.~A.} \bibnamefont{Jim\'enez-Hoyos}},
  \bibinfo{author}{\bibfnamefont{T.~N.} \bibnamefont{Lan}},
  \bibinfo{author}{\bibfnamefont{J.}~\bibnamefont{Li}},
  \bibinfo{author}{\bibfnamefont{F.}~\bibnamefont{Ma}}, \bibnamefont{et~al.},
  \bibinfo{journal}{Phys. Rev. X} \textbf{\bibinfo{volume}{7}},
  \bibinfo{pages}{031059} (\bibinfo{year}{2017}).

\bibitem[{\citenamefont{Stair and
  Evangelista}(2020)}]{stair-evangelista-jcp-2020}
\bibinfo{author}{\bibfnamefont{N.~H.} \bibnamefont{Stair}} \bibnamefont{and}
  \bibinfo{author}{\bibfnamefont{F.~A.} \bibnamefont{Evangelista}},
  \bibinfo{journal}{J. Chem. Phys.} \textbf{\bibinfo{volume}{153}},
  \bibinfo{pages}{104108} (\bibinfo{year}{2020}).

\bibitem[{\citenamefont{Piris}(2019)}]{Piris2019}
\bibinfo{author}{\bibfnamefont{M.}~\bibnamefont{Piris}},
  \bibinfo{journal}{Phys. Rev. A} \textbf{\bibinfo{volume}{100}},
  \bibinfo{pages}{32508} (\bibinfo{year}{2019}).

\bibitem[{\citenamefont{Piris and Mitxelena}(2021)}]{piris2021donof}
\bibinfo{author}{\bibfnamefont{M.}~\bibnamefont{Piris}} \bibnamefont{and}
  \bibinfo{author}{\bibfnamefont{I.}~\bibnamefont{Mitxelena}},
  \bibinfo{journal}{Comput. Phys. Commun} \textbf{\bibinfo{volume}{259}},
  \bibinfo{pages}{107651} (\bibinfo{year}{2021}).

\bibitem[{\citenamefont{Hehre et~al.}(1969)\citenamefont{Hehre, Stewart, and
  Pople}}]{sto-6g}
\bibinfo{author}{\bibfnamefont{W.~J.} \bibnamefont{Hehre}},
  \bibinfo{author}{\bibfnamefont{R.~F.} \bibnamefont{Stewart}},
  \bibnamefont{and} \bibinfo{author}{\bibfnamefont{J.~A.} \bibnamefont{Pople}},
  \bibinfo{journal}{J. Chem. Phys.} \textbf{\bibinfo{volume}{51}},
  \bibinfo{pages}{2657} (\bibinfo{year}{1969}).

\bibitem[{\citenamefont{Hachmann et~al.}(2006)\citenamefont{Hachmann, Cardoen,
  and Chan}}]{hachman-dmrg-2006}
\bibinfo{author}{\bibfnamefont{J.}~\bibnamefont{Hachmann}},
  \bibinfo{author}{\bibfnamefont{W.}~\bibnamefont{Cardoen}}, \bibnamefont{and}
  \bibinfo{author}{\bibfnamefont{G.~K.-L.} \bibnamefont{Chan}},
  \bibinfo{journal}{J. Chem. Phys.} \textbf{\bibinfo{volume}{125}},
  \bibinfo{pages}{144101} (\bibinfo{year}{2006}).

\bibitem[{\citenamefont{Mazziotti}(2011)}]{mazziotti2011H50}
\bibinfo{author}{\bibfnamefont{D.~A.} \bibnamefont{Mazziotti}},
  \bibinfo{journal}{Phys. Rev. Lett.} \textbf{\bibinfo{volume}{106}},
  \bibinfo{pages}{083001} (\bibinfo{year}{2011}).

\bibitem[{\citenamefont{Piris}(2018)}]{piris-2018-dyn}
\bibinfo{author}{\bibfnamefont{M.}~\bibnamefont{Piris}},
  \bibinfo{journal}{Phys. Rev. A} \textbf{\bibinfo{volume}{98}},
  \bibinfo{pages}{022504} (\bibinfo{year}{2018}).

\bibitem[{\citenamefont{Hollett and Loos}(2020)}]{hollet-titou-2020}
\bibinfo{author}{\bibfnamefont{J.~W.} \bibnamefont{Hollett}} \bibnamefont{and}
  \bibinfo{author}{\bibfnamefont{P.-F.} \bibnamefont{Loos}},
  \bibinfo{journal}{J. Chem. Phys.} \textbf{\bibinfo{volume}{152}},
  \bibinfo{pages}{014101} (\bibinfo{year}{2020}).

\bibitem[{\citenamefont{Rodríguez-Mayorga
  et~al.}(2021)\citenamefont{Rodríguez-Mayorga, Mitxelena, Bruneval, and
  Piris}}]{mauricio-2021-jctc}
\bibinfo{author}{\bibfnamefont{M.}~\bibnamefont{Rodríguez-Mayorga}},
  \bibinfo{author}{\bibfnamefont{I.}~\bibnamefont{Mitxelena}},
  \bibinfo{author}{\bibfnamefont{F.}~\bibnamefont{Bruneval}}, \bibnamefont{and}
  \bibinfo{author}{\bibfnamefont{M.}~\bibnamefont{Piris}}, \bibinfo{journal}{J.
  Chem. Theory Comput.} \textbf{\bibinfo{volume}{17}}, \bibinfo{pages}{7562}
  (\bibinfo{year}{2021}).

\bibitem[{\citenamefont{Elayan et~al.}(2022)\citenamefont{Elayan, Gupta, and
  Hollett}}]{elayan-hollet-2022}
\bibinfo{author}{\bibfnamefont{I.~A.} \bibnamefont{Elayan}},
  \bibinfo{author}{\bibfnamefont{R.}~\bibnamefont{Gupta}}, \bibnamefont{and}
  \bibinfo{author}{\bibfnamefont{J.~W.} \bibnamefont{Hollett}},
  \bibinfo{journal}{J. Chem. Phys.} \textbf{\bibinfo{volume}{156}},
  \bibinfo{pages}{094102} (\bibinfo{year}{2022}).

\bibitem[{\citenamefont{Lew-Yee et~al.}(2021)\citenamefont{Lew-Yee, Piris, and
  M.~del Campo}}]{lee-yee-piris-RI}
\bibinfo{author}{\bibfnamefont{J.~F.~H.} \bibnamefont{Lew-Yee}},
  \bibinfo{author}{\bibfnamefont{M.}~\bibnamefont{Piris}}, \bibnamefont{and}
  \bibinfo{author}{\bibfnamefont{J.}~\bibnamefont{M.~del Campo}},
  \bibinfo{journal}{J. Chem. Phys.} \textbf{\bibinfo{volume}{154}},
  \bibinfo{pages}{064102} (\bibinfo{year}{2021}).

\end{thebibliography}
\end{document}